\documentclass[twocolumn,superscriptaddress,longbibliography,aps,prx,floatfix,
preprintnumbers]{revtex4-2}

\usepackage{soul}
\usepackage[dvipsnames]{xcolor}
\usepackage{graphicx}
\usepackage{dcolumn,ulem}
\usepackage{bm}
\usepackage{amsmath}
\usepackage{amssymb}
\usepackage[colorlinks=true,linkcolor=blue,allcolors=blue]{hyperref}%
\usepackage{subfiles}
\usepackage{tabularx}

\begin{document}

\title{Unveiling the interplay of magnetic order and electronic band structure on the evolution of anomalous Hall effect in MnPtGa single crystal}

\author{Gourav~Dwari}
        \email[e-mail: ]{gourav.dwari@tifr.res.in}
		\affiliation{Department of Condensed Matter Physics and Materials Science, Tata Institute of Fundamental Research, Homi Bhabha Road, Colaba, Mumbai 400005, India}

\author{Shovan~Dan}
        \affiliation{Department of Condensed Matter Physics and Materials Science, Tata Institute of Fundamental Research, Homi Bhabha Road, Colaba, Mumbai 400005, India}
        \affiliation{Institute of Low Temperature and Structure Research, Polish Academy of Sciences, Okólna 2, 50-422 Wrocław, Poland}
        
\author{Bishal~Baran~Maity}
		\affiliation{Department of Condensed Matter Physics and Materials Science, Tata Institute of Fundamental Research, Homi Bhabha Road, Colaba, Mumbai 400005, India}

\author{Sitaram~Ramakrishnan}
        \affiliation{Department of Quantum Matter, AdSE, Hiroshima University, 739-8530, Higashi-Hiroshima Japan}
        \affiliation{I-HUB Quantum Technology Foundation, Indian Institute of Science Education and Research, Pune 411008, India}

\author{Achintya~Lakshan}
		\affiliation{Department of Chemistry, IIT Kharagpur, Kharagpur, 721302, India}

\author{Ruta~Kulkarni}
		\affiliation{Department of Condensed Matter Physics and Materials Science, Tata Institute of Fundamental Research, Homi Bhabha Road, Colaba, Mumbai 400005, India}

\author{Vikash~Sharma}
		\affiliation{Department of Condensed Matter Physics and Materials Science, Tata Institute of Fundamental Research, Homi Bhabha Road, Colaba, Mumbai 400005, India}

\author{Suman~Nandi}
		\affiliation{Department of Condensed Matter Physics and Materials Science, Tata Institute of Fundamental Research, Homi Bhabha Road, Colaba, Mumbai 400005, India}
  
\author{Partha~Pratim~Jana}
		\affiliation{Department of Chemistry, IIT Kharagpur, Kharagpur, 721302, India}
  
\author{Andrzej~Ptok}
        \email[e-mail: ]{aptok@mmj.pl}
        \affiliation{Institute of Nuclear Physics, Polish Academy of Sciences, W. E. Radzikowskiego 152, PL-31342 Kraków, Poland}

\author{A.~Thamizhavel}
        \email[e-mail: ]{thamizh@tifr.res.in}
        \affiliation{Department of Condensed Matter Physics and Materials Science, Tata Institute of Fundamental Research, Homi Bhabha Road, Colaba, Mumbai 400005, India}

\date{\today}

\begin{abstract}
The recent studies on the anomalous Hall effect (AHE) have revealed an intrinsic relationship between the topological band structure and the experimentally observed transverse conductivity. Consequently, this has led to a heightened focus on examining the topological aspects of AHE. Here we have studied sign reversal of anomalous Hall conductivity with temperature in the single crystalline MnPtGa (space group: $P6_3/mmc$). From the interdependence of the linear resistance, we claim that the origin of such behavior is intrinsic. By systematically studying the electronic band structure and Berry curvature of MnPtGa using first principle calculations supported by magnetic susceptibility and isothermal magnetization measurements we demonstrate that the temperature dependent complex magnetic structure plays a significant role and leads to the sign reversal of anomalous Hall conductivity. We proposed a continuous evolution of the magnetic structure, supported by the ab initio calculations, which is consistent with the experimental data. Our studies have established that the critical temperature ($\approx$110 K), where the sign reversal appears is associated with the magnetic structure and the magnitude of Mn moments.
\end{abstract}

\maketitle

\section*{Introduction}

Although identified long ago, the complete understanding of the anomalous Hall effect (AHE), involving time-reversal symmetry breaking, is yet to be resolved. Primarily, it was argued to originate from intrinsic or extrinsic contributions~\cite{Nagaosa.Sinova.2010}. Intrinsic AHE originates from the magnetism, band topology, Berry curvature etc.~\cite{Jungwirth.Niu.2002, Onoda.Nagaosa.2003, Yao.Kleinman.2004}. Extrinsic AHE, on the other hand, originates from scattering mechanisms like skew scattering and side jump due to spin-orbit interaction~\cite{Smit.1955, Smit.1958, Berger.1970}. {Despite its complex origins, the anomalous Hall resistivity ($\rho_{yx}^A$) exhibits a striking association with the linear resistivity ($\rho_{xx}$) through a power-law relationship, $\rho_{yx}^A\propto \rho_{xx}^{\beta}$ . This interconnection holds crucial implications for understanding the underlying mechanisms of the AHE}
~\cite{Nagaosa.Sinova.2010, Karplus.Luttinger.1954}.
Depending on the values of the exponent $\beta$, it was argued to originate from the intrinsic or extrinsic mechanism. Nevertheless, a subsequent theoretical model was proposed to support the experimental observation of the AHE. 
Moreover, the specific topological band involved in the electronic structure plays a crucial role in determining both the magnitude and sign of the anomalous Hall conductance (AHC) in the intrinsic AHE. 
Depending on the position of the band, or the modification of the band by the external parameters, the values of the AHC may change by a significant amount. Such phenomena were observed in a series of metallic compounds, viz., Co$_3$Sn$_2$S$_2$~\cite{Wang.Xu.2018,Liu.Sun.2018}, Co$_2$MnGa~\cite{Zhang.Yin.2021}, and Fe$_3$GeTe$_2$~\cite{kim.Seo.2018}.
An alteration in the sign was noted in a Weyl semimetal, HoPtBi, and it was suggested that this was caused by the splitting of Weyl nodes, which is modulated by exchange splitting~\cite{Chen.Xu.2021}. 
\begin{figure*}%[!t]
	\includegraphics[width=0.9\textwidth]{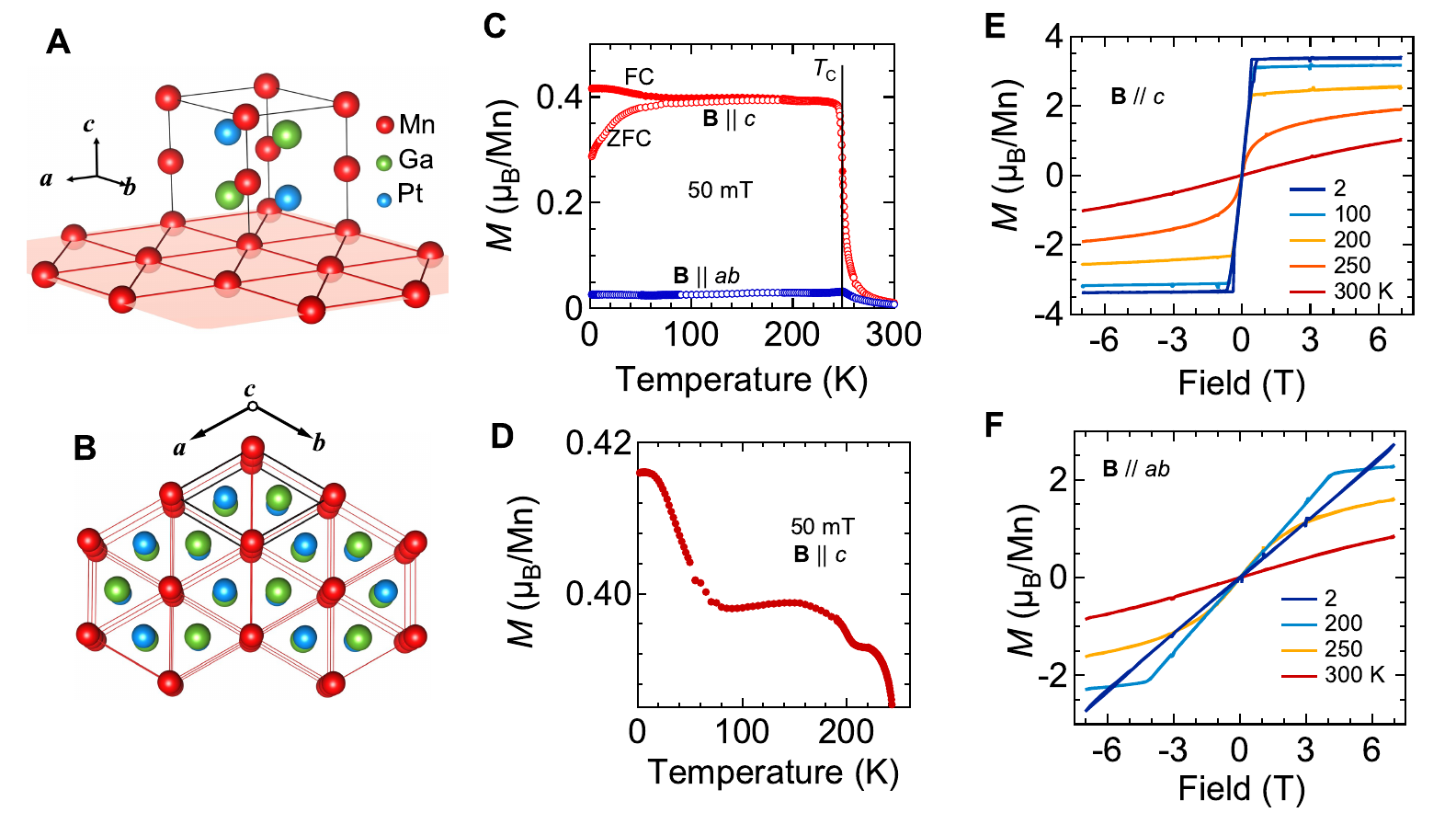}
	\caption{Structure and magnetization of MnPtGa. (\textbf{A}),(\textbf{B}) Centrosymmetric crystal structure of MnPtGa where red, green, and blue balls represent the Mn, Ga, and Pt atoms, respectively. (\textbf{C}) Temperature dependence of magnetization $M$ with FC and ZFC protocols for 50~mT magnetic field applied along $c$-direction and in $ab$-plane. (\textbf{D}) Enlarged $M$ vs. $T$ for \textbf{B} in $c$-direction with FC protocol, depicting multiple magnetic transitions. Magnetic field dependence of $M$ at selected temperatures $B\|c$ and $B\|ab$-plane are depicted in (\textbf{E}) and (\textbf{F}) respectively.}
	\label{Fig1}
\end{figure*}
The sign change in AHC has also been observed in epitaxial thin films of  La-doped EuTiO$_3$, irrespecitve of the synthesis conditions~\cite{Takahashi.Ishizuka.2018}. 
In that case, the change of sign was attributed to the chemical potential shift (as a function of the magnetic field). 
Probably, the most interesting observation of the AHE was reported in the itinerant ferromagnet SrRuO$_3$, where the variation of the AHC was observed as a function of temperature. 
A correlation between structure and magnetism was discussed to draw a link with the magnetic monopole~\cite{mathieu.asamitsu.04}. A similar discussion on the monopole and the Berry curvature was discussed for the compound Co$_3$Sn$_2$S$_2$~\cite{Liu.Sun.2018}. 
Reports also suggest that it is possible to tune the AHE in compounds by doping, so that the Fermi level lies between the overlap of two narrow bands which leads to a change in sign of effective spin-orbit parameter resulting in a sign reversal of AHC as observed in semiconducting family BaFe$_{2\pm x}$Ru$_{4\mp x}$O$_{11}$~\cite{Shlyk.Niewa.2010} and metallic spinel CuCr$_2$Se$_{4-x}$Br$_x$~\cite{Lee.Watauchi.2004}. However, such mechanism does not cause a non-monotonicity in AHC.

The AHE in thin films of MnPtGa was recently reported to exhibit a strict non-monotonic behavior and it was claimed that the sign reversal ($\sim110~$K) of AHC is very robust with temperature,  and the variation in thickness does not affect this phenomenon~\cite{Ibarra.Lesne.Sabir.2022}. 
Although, the authors of Ref.~\cite{Ibarra.Lesne.Sabir.2022} performed a temperature-dependent neutron diffraction (ND) study of the compound~\cite{Ibarra.Lesne.2022}, the justification for the robustness of the temperature at which the sign of AHC reverses still remains unsolved.
For the same compound, in polycrystalline form, a serious ND study was performed, notwithstanding the focus was different~\cite{Cooley.Bocarsly.2020}. It is worth mentioning here, that the compound was earlier reported to show polymorphism with different crystal structures, viz., cubic (space group $F\bar{4}3m$)~\cite{Hames.Crangle.1971, Dunlap.Jha.1983}, trigonal (space group $P3m1$)~\cite{Srivastava.Devi.2020} and hexagonal (space group $P6_3/mmc$)~\cite{Cooley.Bocarsly.2020, Ibarra.Lesne.2022}. 
Among these, the trigonal structure is non-centrosymmetric and hosts stable N\'eel-type skyrmion over a wide temperature range~\cite{Srivastava.Devi.2020}. 
Studies on the hexagonal polymorphs at low temperatures uncover the realization of the long-range magnetic order which can lead to less symmetric structure, like $Amm2$ or $C2/m$~\cite{Cooley.Bocarsly.2020}. In the present study, we revisited the AHE of high-quality single crystals of MnPtGa that crystallized in the hexagonal structure. We have used the \textit{ab initio} calculations to correlate the sign reversal of the AHC with the electronic band structure modification due to the temperature dependence of the magnetic order.

\section*{Results and Discussions}
 \begin{figure*}[!ht]
	\includegraphics[width=0.9\linewidth]{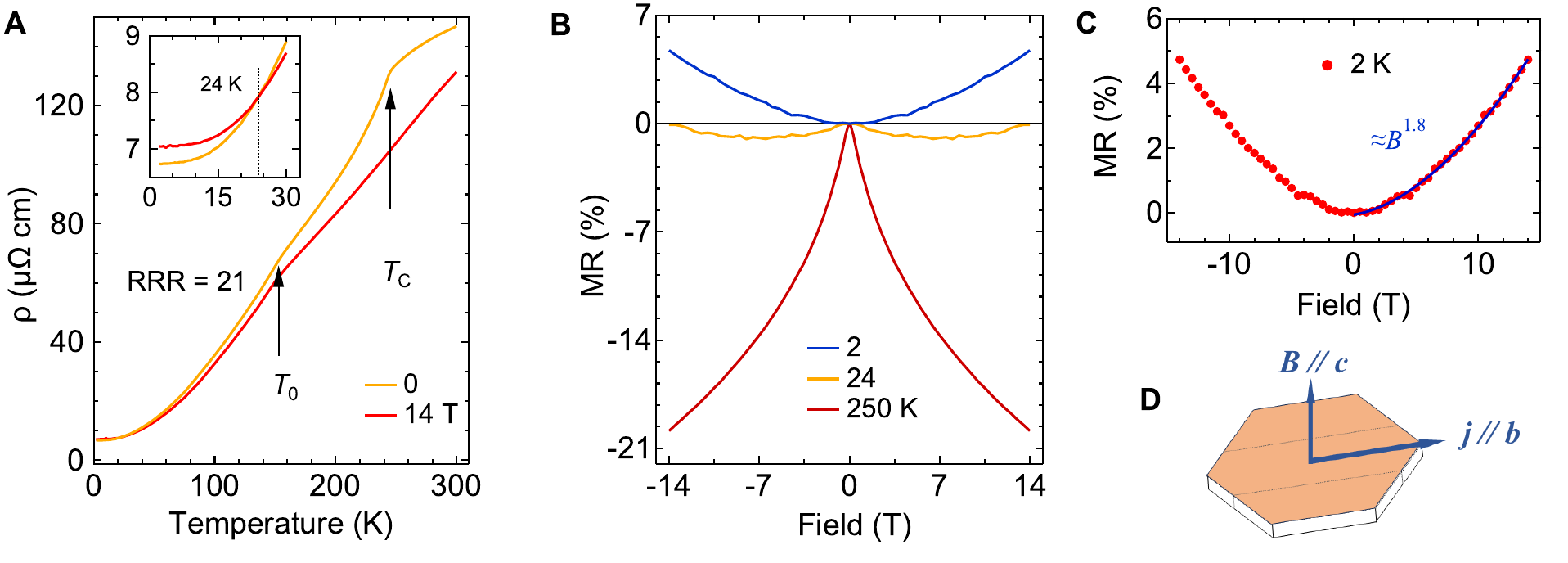}
	\caption{(\textbf{A}) Linear Resistivity $\rho$ versus temperature of MnPtGa single crystal in $0$ and $14$~T magnetic field applied along c-axis (left axis) and temperature derivative of $\rho$ at $0$~T (right axis) indicating the ferromagnetic Curie temperature $T_{\rm C}$ and $T_{0}$ at which the canting of the spins occur. Inset shows a crossover point around $24$~K below which the magnetoresistance becomes negative to positive at $14$~T. (\textbf{B}) Magnetic field dependence of magnetoresistance, MR, at three selected temperatures of $2$, $24$ and $250$~K. (\textbf{C}) Positive MR of MnPtGa at $2$~K fitted with a power law with exponent $1.8$ indicating semimetallic nature. (\textbf{D}) Schematic of the current and applied magnetic field used to measure linear resistivity. }
	\label{Fig2} 
\end{figure*}

\subsection*{Structural Analysis}
 
As MnPtGa shows polymorphism with three different crystal structures, it is crucial to determine that in which polymorphic form our Bridgman-grown MnPtGa has crystallized into~\cite{Hames.Crangle.1971,Dunlap.Jha.1983,Srivastava.Devi.2020,Cooley.Bocarsly.2020,Ibarra.Lesne.2022}.
We have used powder x-ray diffraction and the Laue technique, shown in Fig.~\ref{XRD}(a),(b), as a preliminary tool of identification that clearly rules out the cubic structure. It is to be noted that polycrystalline material synthesized by arc melting is reported to adopt the Ni$_2$In structure type described by the space group $P6_3/mmc$ as shown in Fig.~\ref{Fig1}(a) and~\ref{Fig1}(b), where its characteristic feature is a honeycomb lattice involving a stacking of Mn with Pt-Ga atoms along the $c$-direction (ABAB fashion)~\cite{Cooley.Bocarsly.2020}. However, in one of the recent reports, the single crystals grown by Bridgman method was reported to crystallize in $P3m1$ trigonal space group breaking the six-fold symmetry~\cite{Srivastava.Devi.2020}. This results in an acentric structure defined by the polar space group $P3m1$, which is essentially a sub-group of $P6_3/mmc$ following the B\"{a}rnighausen formalism. With the acentric structure, there are 3 additional atoms:  Mn, Pt and Ga split into two, where the site symmetry is also $3m$ for the new atoms (See Table \ref{compare_ref}). However, single crystal x-ray diffraction (SXRD) analysis (see \textit{Methods}) on our sample suggests that the center of inversion is still present in the crystal structure and thus can be described by $P6_3/mmc$. This is reinforced by no improvement to the model by refinement of additional parameters for the acentric $P3m1$. See Tables \ref{compare_model} and \ref{compare_ref} for details.

\subsection*{Magnetic Properties} 

Measurement of magnetization as a function of temperature ($M(T)$) in $0.05$~T magnetic field (Fig.~\ref{Fig1}(d)) applied both in-plane and out-of-plane directions, demonstrates that MnPtGa undergoes a ferromagnetic transition at $T_{\rm C}=250$~K.
However, when the temperature is further lowered MnPtGa enters into a spin-canted state at around $\sim165$~K ($T_{0}$) indicated by the broad hump for field along both the principal crystallographic directions. 
From powder neutron diffraction experiments, the spin-canted state can be understood as the resultant of two components: a ferromagnetic component along the $c$-direction together with an antiferromagnetic moment pointed in the $a$-direction~\cite{Cooley.Bocarsly.2020}.
Additional kinks in $M(T)$ when lowering the temperature suggest a rich magnetic structure that evolves with temperature. 
Indeed, at lower temperatures ($< 150$~K), the magnetic order changes from spin-canted to canted antiferromagnetic spin density wave with moments oriented in the hexagonal $c$ direction and flipping basal plane~\cite{Cooley.Bocarsly.2020}.
Moreover, magnetostriction studies~\cite{Cooley.Bocarsly.2020} have shown that these transitions are coupled with the lattice and lead to negative volume magnetostriction in MnPtGa.
Our field-dependent magnetization ($M(B)$) measurements suggest that MnPtGa is a soft ferromagnet with strong magnetocrystalline anisotropy.
The saturation field ($\sim 0.5~\rm T$) is fairly low along the $c$-axis and extremely large ($\sim 8.5~\rm T$) in the $ab-$plane, at 2~K, confirming dominant out-of-plane magnetization. 
However, unlike the thin film and polycrystalline MnPtGa~\cite{Cooley.Bocarsly.2020, Ibarra.Lesne.Sabir.2022}, we found negligible coercivity in the single crystalline form indicating a rather soft ferromagnetic nature. 
Experimentally observed saturation magnetization is almost constant in a wide temperature range ($\sim 2-50$~K) with a value of $3.3$~$\mu_{\rm B}$/Mn which is very close to the theoretical moment of $3.6$~$\mu_{\rm B}$/Mn, implying a complete ferromagnetic state is achieved by applying $\sim 0.5$~T magnetic filed in the out-of-plane direction. 

Additionally, we performed the theoretical calculations for different Mn-magnetic moment orientations (i.e. magnetic moments along $c$, and tilted from $c$~axis)~\cite{Ibarra.Lesne.Sabir.2022, Ibarra.Lesne.2022}.
These findings support previous studies of the magnetic order in MnPtGa~\cite{Ibarra.Lesne.Sabir.2022,Ibarra.Lesne.2022}.
The calculated magnetic anisotropy energy (MAE) is around $0.9$~meV/f.u., while the estimated magnetic moment is equals to $3.6$~$\mu_{\rm B}$.
The magnetic moment value is close to this reported in previous theoretical study~\cite{Cooley.Bocarsly.2020}, however larger than observed experimentally (i.e. around $3.3$~$\mu_{\rm B}$ in our study, and in Refs.~\cite{buschow.engen.83,Cooley.Bocarsly.2020}).
The observed difference can be related to realization long-range magnetic order at low temperatures, due to the combination of canted-spin order and spin density wave~\cite{Cooley.Bocarsly.2020}.
Moreover, from optimization of the crystal structure with different magnetic moment, we found that the emergence of non-collinear magnetic order (with tilted magnetic state) can lead to the local symmetry breaking, and lowering of the crystal structure.
Nevertheless, such ``local'' symmetry breaking does not affect the system stability (see Sec.~\ref{sec.sm_theo} in Supplemental Information), while it can affect the global electronic properties of the system with long-range magnetic order~\cite{Cooley.Bocarsly.2020}.

\begin{figure*}[!ht]
	\includegraphics[width=0.9\textwidth]{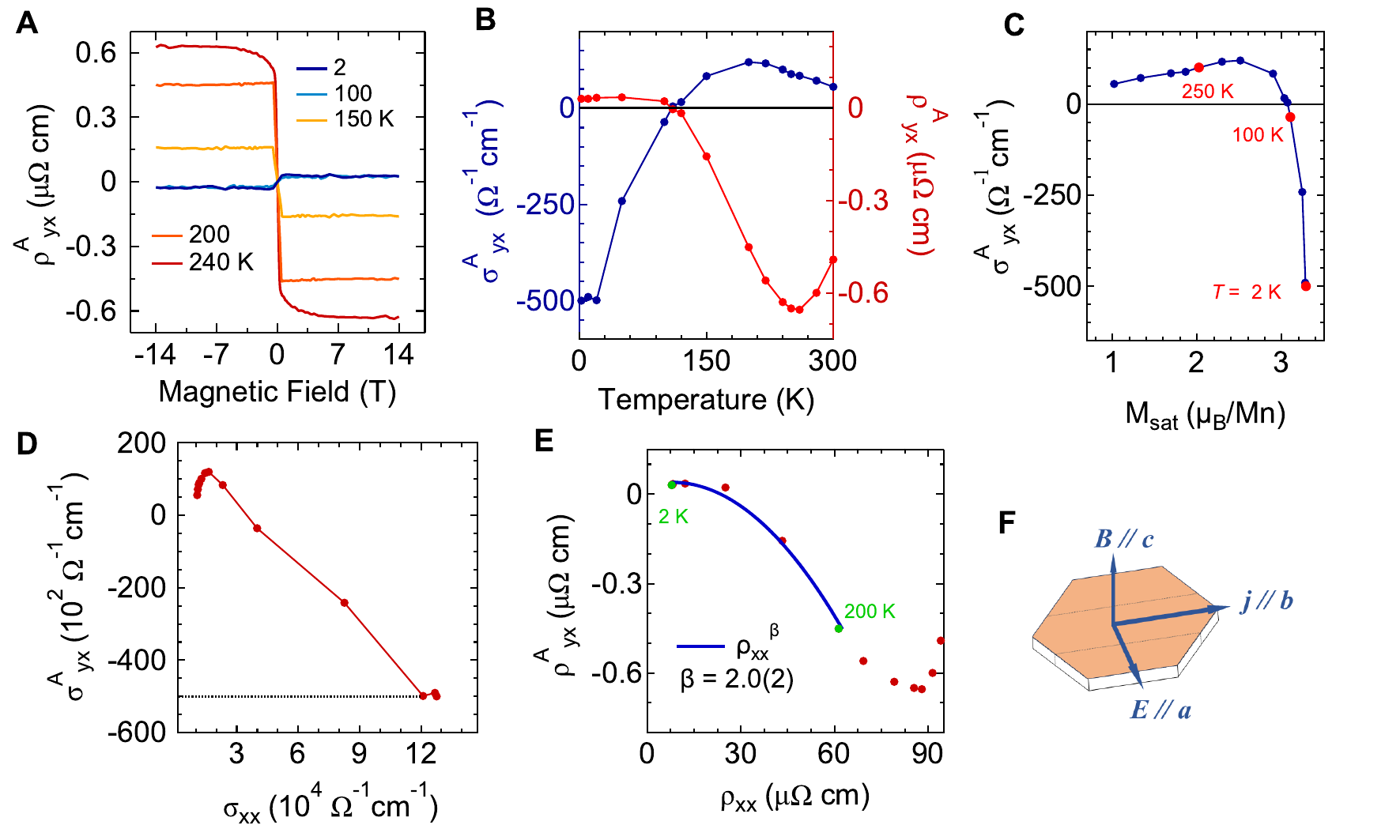}
	\caption{Hall effect in MnPtGa. (\textbf{A}) $\rho_{yx}^A$ as a function of the $B$ at different temperatures. (\textbf{B}) $T$ dependence of $\sigma_{yx}^A$ (left axis) and $\rho_{yx}^A$ (right axis) at zero magnetic fields, depicting the sign reversal of AHE. (\textbf{C}) Plot of $\sigma_{yx}^A$ vs. saturated magnetization $M_{sat}$.(\textbf{D}) $\sigma_{xx}$ dependence of $\sigma_{yx}^A$. $\sigma_{yx}^A$ becomes independent of $\sigma_{xx}$ below 20~K temperature. (\textbf{E}) Plot of $\rho_{yx}^A$ vs. $\rho_{xx}$ with temperature as a parameter. The solid line represents the fitting with $\rho_{yx}^A\propto (\rho_{xx})^{\beta}$ relation. (\textbf{F}) Schematic diagram of sample geometry along with current and applied magnetic field directions.
    }
	\label{Fig3} 
 \end{figure*}

\subsection*{Linear and Hall resistivity}

As evident from the longitudinal electrical resistivity ($\rho$) of MnPtGa, shown in Fig.~\ref{Fig2}(a), we have grown a high-quality single crystal with a residual resistivity ratio (RRR) of $\thickapprox$ 22 reflecting a coherent growth and good quality of the single crystal. MnPtGa exhibits a metal-like behavior where the resistivity decreases with decreasing temperature with a sharp drop below $T_{\rm C}$ due to a reduction in spin-disordered scattering. Additionally, a broad hump around $T_{0}$ indicates the onset of the canted magnetic state. As the field is ramped up to 14~T, a negative magnetoresistance with large value at $T_{\rm C}$, is observed over a large temperature range that is typical for ferromagnetic metals. But the hump around $T_{0}$ persists even at 14~T. This indicates that the hump may be attributed to the structural transition, however, due to magneto-structural coupling, a structural deformation leads to a canted magnetic ground state at zero magnetic field in MnPtGa near $T_0$, which is also observed and argued to have the similar origin from the magnetostriction analysis in the previous study~\cite{Cooley.Bocarsly.2020}.

When the temperature is further lowered, the negative magnetoresistance becomes progressively smaller, and a crossover is observed where the magnetoresistance changes sign and becomes positive below 24~K, shown in the inset of Fig.~\ref{Fig2}(a). This is also evident from Fig.~\ref{Fig2}(b) where field-dependent magnetoresistance is plotted. At 24~K, the MR has distinct features: near zero field, MR decreases with increasing magnetic field but beyond 7~T it increases unboundedly, indicating the crossover temperature is not constant but varies with the applied magnetic field. At 2~K MR is completely positive and increases without any saturation up to a magnetic field of 14~T. An attempt to fit the MR data at 2~K with $MR\sim B^n$ yielded an exponent value of 1.8, indicating a typical semimetallic nature. To further investigate this behavior, we have used the semiclassical two-band model to fit $\rho_{xx}(B)$ and $\rho_{yx}(B)$ simultaneously for fields above 3~T where the anomalous term ($c$) in $\rho_{yx}(B)$ can be considered as constant:
\begin{equation}
    \rho_{yx}=c+\frac{1}{e}\left[\frac{(n_h\mu_h^2-n_e\mu_e^2)B+(n_h-n_e)\mu_h^2\mu_e^2B^3}{(n_h\mu_h+n_e\mu_e^2)^2+(n_h-n_e)^2\mu_h^2\mu_e^2B^2}\right],
    \label{eq1}
\end{equation}
 Extracted mobilities are $\mu_e=1.61(8)\times~10^{2}$~cm$^2$V$^{-1}$s$^{-1}$ and $\mu_h=1.48(7)\times~10^{2}$~cm$^2$V$^{-1}$s$^{-1}$ for electrons and holes, with carrier densities $n_e=2.61(9)\times~10^{21}$~cm$^{-3}$ and $n_e=3.40(6)\times~10^{21}$~cm$^{-3}$ respectively, which is one order of magnitude smaller than the reported carrier density in MnPtGa thin film. However, estimated carrier densities are slightly larger than the semimetallic regime and close to each other. Small values of mobilities also justify the negligible magnetoresistance at 2~K.
 
\begin{figure*}[!ht]
    \includegraphics[width=\linewidth]{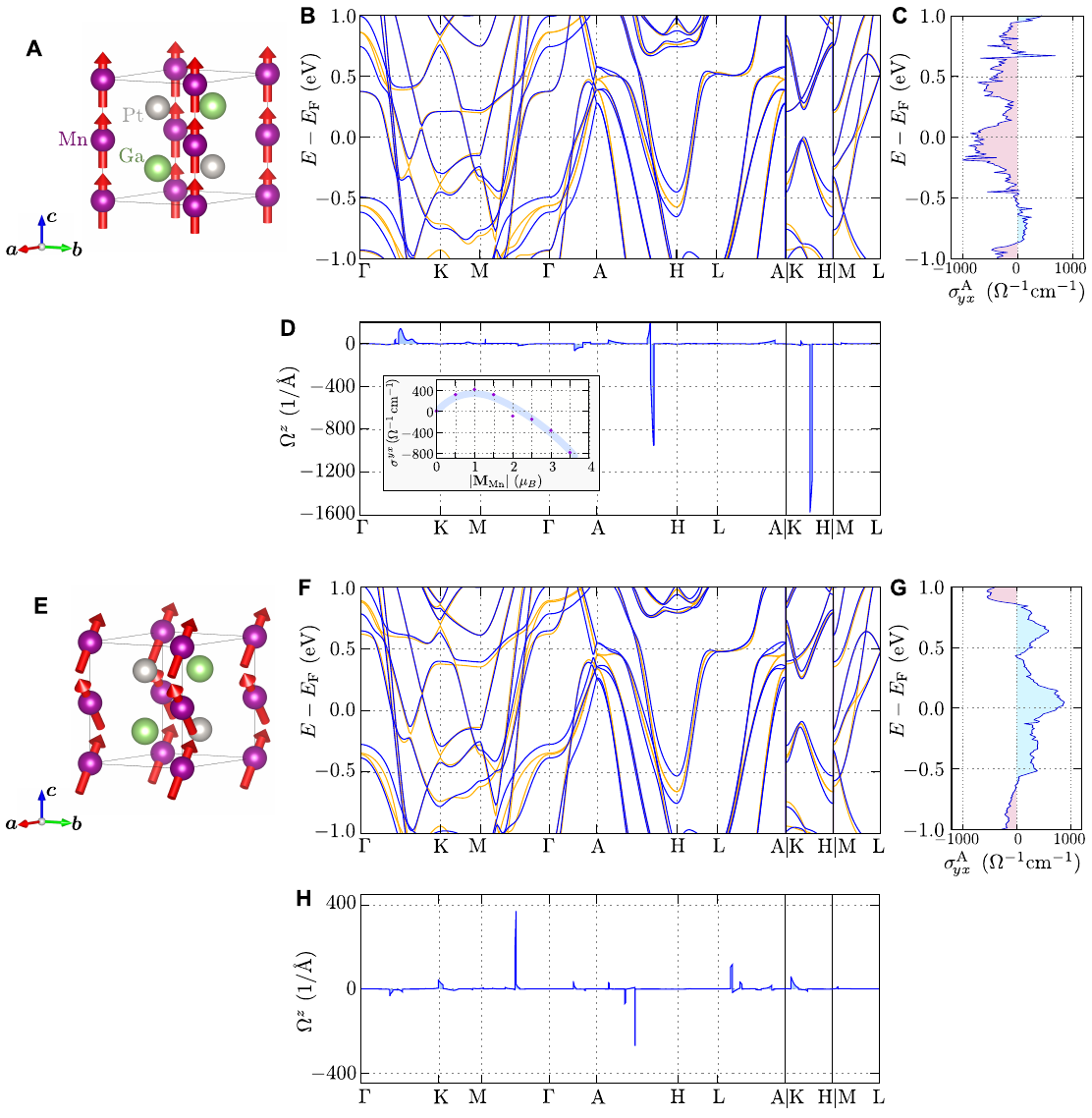}
	\caption{
    The influence of the magnetic order on the electronic transport properties of MnPtGa.
    Top panels (\textbf{B})--(\textbf{D}) and bottom panels (\textbf{F})--(\textbf{H}) present results for the Mn magnetic moment along $c$-direction (\textbf{A}) and tilted from $c$-direction (\textbf{E}), respectively.
    The electronic band structure (\textbf{B},\textbf{F}), the Berry curvature $\Omega_{z}(k)$ on the Fermi level along high symmetry directions (\textbf{C},\textbf{G}), and the Fermi level dependence of AHE $\sigma^{A}_{yx}$ (\textbf{D},\textbf{H}).
    Inset on (\textbf{D}), present magnetic moment dependence of AHE $\sigma^{A}_{yx}$.}
	\label{fig.theo}
 
\end{figure*}

Our transport measurements reveal a robust signature of the anomalous Hall effect for out-of-plane configuration of $j\| b$ and $B\| c$ but vanish for in-plane field directions. Total Hall resistivity ($\rho_{yx}$) of MnPtGa can be decomposed into two parts:
\begin{equation}
     \rho_{yx}=R_0B+\rho_{yx}^{A}
     \label{eqn:Hall}
\end{equation}
where the first term denotes the ordinary Hall contribution ($\rho_{yx}^{O}$) due to Lorentz force and $\rho_{yx}^{A}$ is the anomalous Hall resistivity. As anomalous Hall resistivity is proportional to the magnetization M of the sample, $\rho_{yx}^{A}$ becomes also constant beyond the saturation of \textit{M}, thus a linear fitting will produce the intercept as anomalous Hall term with the slope as $R_0$. The linear behavior of normal Hall can be justified from Eq.~(\ref{eq1}), where for $n_e\cong n_h$, the $B^3$ and $B^2$ terms from numerator and denominator become negligible and $\rho_{yx}$ reduces to linear in \textit{B}, that is observed in MnPtGa above the field where \textit{M} is saturated. 
From the plot of the magnetic field variation of $\rho_{yx}^A$, in Fig.~\ref{Fig3}(a), we observe that $\rho_{yx}^A$ is negative around $T_{\rm N}$ but becomes positive when lowering the temperature. Temperature dependence of $\sigma_{yx}^A$, Fig.~\ref{Fig3}(b), depicts the sign change more clearly near the temperature 109~K ($T^*$), which has also been seen in the thin film of MnPtGa. \cite{Ibarra.Lesne.Sabir.2022}.

The observed anomalous Hall effect can be dominated either by extrinsic impurity scattering or by intrinsic mechanisms. To investigate the underlying mechanism, we have fitted $\rho_{yx}^A$ data with $\rho_{xx}$, which exhibits quadratic behavior as shown in Fig.~\ref{Fig3}(E).. This rules out the presence of the skew-scattering mechanism, which is well known to vary linearly with $\rho_{xx}$. The observed quadratic behavior suggest that the extrinsic side-jump or intrinsic Berry curvature mechanism may be responsible for the observed AHE in MnPtGa~\cite{Ibarra.Lesne.2022}.

AHC for the side-jump contribution in multiband ferromagnets with dilute impurity can be estimated from the expression, $(e^2/(ha))\cdot(\epsilon_{SO}/E_{\rm F})$, where $\epsilon_{SO}$ is the spin-orbit interaction energy, and $a$ is the lattice constant~\cite{Onoda.Sugimoto.2006}. Using the lattice constant, $a\sim V^{1/3}=4.495$~\AA\ 
and estimated $\epsilon_{SO}/E_{\rm F}\sim0.01$, the derived side-jump contribution is $8.6~\Omega^{-1}$cm$^{-1}$, which is very small compared to the observed AHC.
A similar value has been obtained in ferromagnetic half metal Co$_3$Sn$_2$S$_2$ for side-jump contribution~\cite{Wang.Xu.2018}.
Moreover, the longitudinal conductivity ($\sigma_{xx}$) of MnPtGa lies in the intermediate regime ($2\times10^{3}<\sigma_{xx}<6\times10^{5}~\Omega ^{-1}$cm$^{-1}$) which indicates AHC of MnPtGa is neither dominated by skew scattering (high conductivity regime, $\sigma_{xx}>6\times10^{5}~\Omega ^{-1}$cm$^{-1}$) nor by impurity scattering (bad metal hopping region, $\sigma_{xx}<2\times10^{3}~\Omega ^{-1}$cm$^{-1}$).
However, our observation from the temperature dependence of the $\sigma_{yx}^A$ plot suggests that below 20~K temperature (below the crossover temperature, in the semimetallic state), $\sigma_{yx}^A$ becomes nearly constant.
This behavior of $\sigma_{yx}^A$ is also observed in the anomalous Hall compound Co$_3$Sn$_2$S$_2$ (for example) in a wider range of temperature below 100~K which agrees with the prediction of the unified model for AHE physics for the intrinsic anomalous Hall effect~\cite{Liu.Sun.2018}. Collectively, these indicate that the observed anomalous Hall effect in MnPtGa is originating from the intrinsic mechanism dominated by the Berry curvature.

\subsection*{Electronic Band Structure}

We have performed a detailed electronic structure study to investigate the proposed mechanism of AHE based on experimental results.
Theoretical calculations, assuming full optimization of the system, uncover the ground state with the magnetic moment tilted from the $c$-direction.
However, tilting of the magnetic moment leads to the breaking of the symmetry resulting in a lower symmetry $P2_1/m$ (see Sec.~\ref{sec.sm_theo} in Supplemental Information).
Contrary to this, the Mn magnetic moments along $c$-direction preserved the $P6_3/mmc$ symmetry.
Both structures are stable in a dynamic sense and can be used to mimic ``true'' structures at different temperatures.
As mentioned earlier, the experimental investigation related to the low-temperature long-range magnetic order~\cite{Cooley.Bocarsly.2020}, suggests the realization of a magnetic unit cell with symmetry $C2/m$ (around $10$~K) or $Amm2$ (around 50~K).
Indeed, the low-temperature magnetic unit cell can be understood as a combination of both magnetic orders (with magnetic moment along $c$-direction and tilted from $c$-direction).
Accordingly, results for systems with Mn magnetic moments along the $c-$ direction and tilted from the $c-$direction can be used to explain the reported results (Fig.~\ref{fig.theo}).
\begin{figure*}%[!h]
    \centering
    \includegraphics[width=0.8\textwidth]{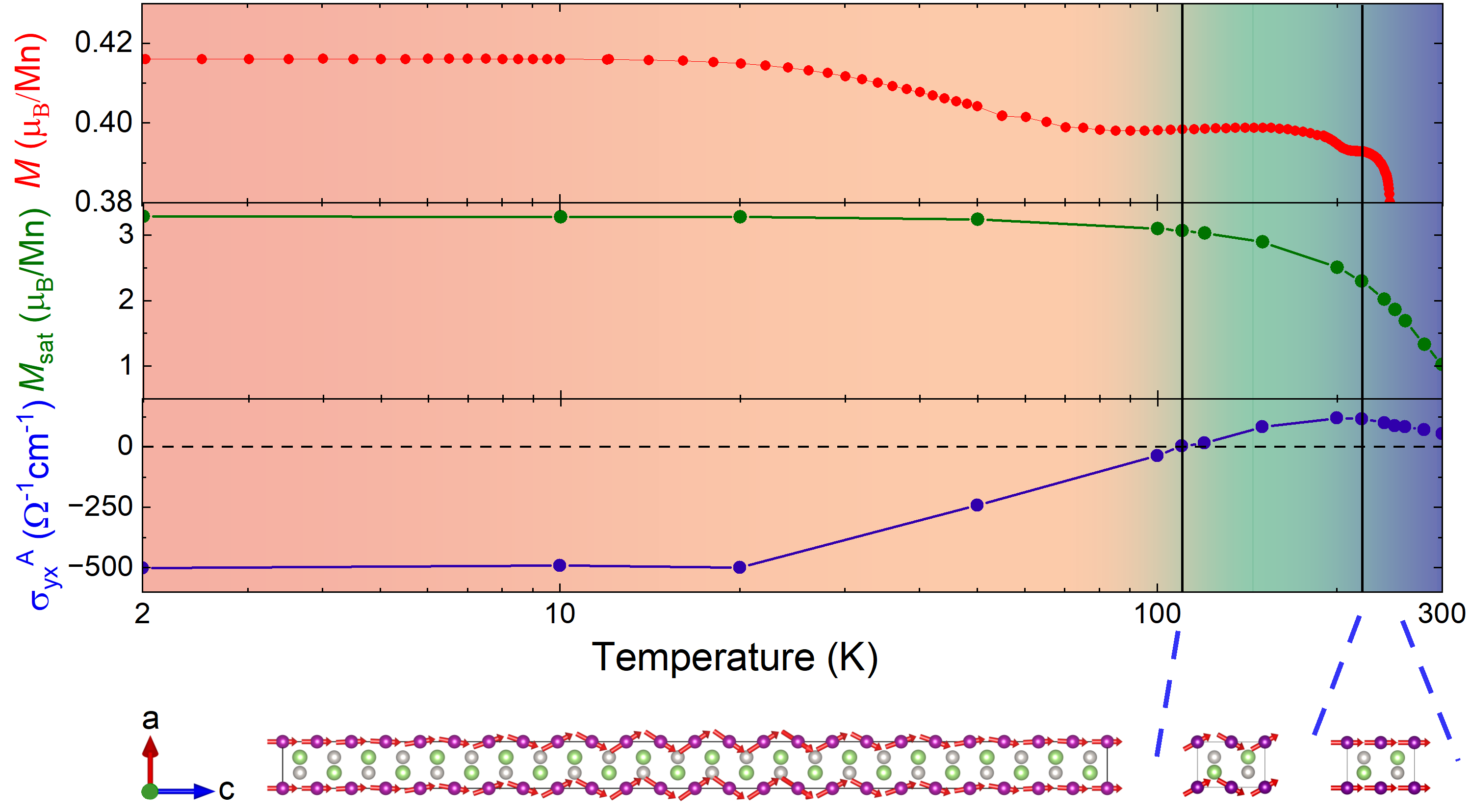}
    \caption{(from top) The evolution of magnetization, anomalous Hall conductivity, and saturation magnetization of MnPtGa with temperature. Cartoons of different magnetic structures at corresponding temperature ranges are also shown at the bottom panel.}
    \label{Evolution}
\end{figure*}

The electronic band structure in the absence and presence of the spin-orbit coupling for the magnetic moments along $c$-direction and tilted from $c$-direction are shown in Fig.~\ref{fig.theo}(A) and (D), respectively. 
In the absence of spin-orbit coupling, several band crossings can be found in the vicinity of the Fermi level for both magnetic orders.
Introducing this spin-orbit coupling leads to the opening of well-visible band gaps in the electronic band structure (cf. orange and blue lines on Fig.~\ref{fig.theo}(D) and~\ref{fig.theo}(H), see also Fig.~\ref{fig.colinear} in Supplemental Information).
Arising spin-mixing states around this gap contribute to the large Berry curvature~\cite{yao.kleinman.04, wang.yates.06}:
\begin{eqnarray}
	\Omega_{n}^{z} ( {\bm k} ) = - 2 \text{Im} \sum_{m \neq n} \frac{ \langle \Psi_{n{\bm k}} \vert v_{x} \vert \Psi_{m{\bm k}} \rangle \langle \Psi_{m{\bm k}} \vert v_{y} \vert \Psi_{n{\bm k}} \rangle }{ ( E_{m{\bm k}} - E_{n{\bm k}} )^{2} } ,
\end{eqnarray}
where $v_{x(y)}$ is the velocity operator, while $E_{n{\bm k}}$ and $\langle \Psi_{n{\bm k}}$ eigenpair of $n$th band at momentum ${\bm k}$.
When $E_{m{\bm k}}$ is close to $E_{n{\bm k}}$, then large Berry curvature emerge.
Indeed, this is clear in Fig.~\ref{fig.theo}(B) and Fig.~\ref{fig.theo}(E), where the $z$-component of the Berry curvature $\Omega^{z}$ at the Fermi level is presented.
It is important to highlight that the alteration of the magnetic order, achieved by tilting the Mn magnetic moments away from the $c$-direction, resulted in a substantial modification of the obtained results, as illustrated in Fig.~\ref{fig.theo}(F),(G) and (H).

The non-vanishing total Berry curvature over the Brillouin zone gives rise to the AHE with $\sigma^A_{yx} \sim \int_{\text{BZ}} d^{3}{\bm k} \sum_{n} f_{n} ( E_{n{\bm k}} ) \Omega_{z}^{n} ( {\bm k} )$, where $f_{n} ( E )$ is the Fermi-Dirac distribution, and $n$ is the index of the occupied bands (see Fig.~\ref{fig.theo}(C) and~\ref{fig.theo}(G)).
Here we would like to mention, that the electronic properties strongly depend on the Mn magnetic moment direction.
In particular, tilting of the Mn magnetic moment from the $c$-direction leads to the change of the AHE sign, [cf.~Fig.~\ref{fig.theo}(c) and (g)].
The link between theoretically obtained structures and experimentally observed magnetic order uncovers a mechanism leading to the change of sign in MnPtGa with temperature (presented in Fig.~\ref{Fig3}(c)).
For example, at the high-temperature phase, when the magnetic moment is titled from the $c$-direction~\cite{Cooley.Bocarsly.2020}, the AHC is positive.
Conversely, decreasing the temperature results in the stabilization of long-range magnetic order, particularly affecting the phase with magnetic order along the $c$-direction (illustrated by negative AHE in Fig.~\ref{Fig3}(c)).
From this, decreasing temperature leads to the changes in AHE sign to negative (Fig.~\ref{Evolution}).
A similar effect was reported in Sr$_{1-x}$Ca$_x$RuO$_3$ series ($0\leqslant x\leqslant 0.4$)~\cite{mathieu.asamitsu.04}, where the AHE changed sign with temperature during the modification of magnetic moment.
In this context, the present situation becomes more intricate as lowering the temperature results in both modifications to the magnetic order and a change in the sign of AHE.
Indeed, the AHE sign change is observed around temperature, where the low-temperature long-range magnetic order emerges.

\section*{Summary}

One of the less explored and most exciting phenomenon in the field of AHE is probably the sign reversal of AHC, which we have studied in this article in the spin-canted magnet MnPtGa using magneto-transport measurements and theoretical investigations. The scaling behavior of $\rho_{yx}^A$ with $\rho_{xx}$ together with the unified model of AHE, we have concluded that the observed AHE in MnPtGa is originating from the momentum space Berry curvature, as per the existing reports. A complex magnetic structure has been substantiated by our magnetization measurements. 
We have estimated the temperature dependence of $\sigma_{yx}^A$ from first principle calculation by considering the temperature dependence of the Bloch state exchange splitting which is proportional to the temperature evolution of magnetization.
To voice our experimentally observed sign reversal of $\sigma_{yx}^A$,  we have used $ab$-$initio$ calculations, $M(T)$ data, and magnetic structures determined from neutron diffraction (from ref:\cite{Cooley.Bocarsly.2020} and \cite{Ibarra.Lesne.2022}). For easy understanding, we have split the complete temperature range into three slices (Fig.~\ref{Evolution}): 
(i) for $T>220$~K, the system is in the ferromagnetic state with relatively low magnitude of magnetic moments, aligned along $c$-axis. In this temperature range, experimental $\sigma_{yx}^A$ is positive. Our calculated $\sigma_{yx}^A$ (see inset of Fig.\ref{fig.theo}(d)) also depicts a positive value when magnetization goes below $\sim$2~$\mu_{\rm B}$ which corresponds to near $T_{\rm C}$.
(ii) In 110~K$<T<$220~K range, the canted magnetic state becomes more stable, which is evident from neutron diffraction and the downturn of $M(T)$ data as the temperature is lowered. The calculated AHC also exhibits a positive value in the canted magnetic state, which follows our experimental results.
(iii) As the temperature is further lowered, there is competition between ferromagnetic and canted magnetic interactions. However, our $M(T)$ data exhibit a ferromagnetic type upturn indicating a dominant ferromagnetic interaction, which leads to a negative sign of $\sigma_{yx}^A$. In Ref.~\onlinecite{Cooley.Bocarsly.2020,Ibarra.Lesne.2022}, it was established from the ND study, that the magnetic ground state at low temperatures is stabilized with a combined canted AFM and spin density wave structure, with a large magnitude of the magnetic moment among crystallographic $c-$direction. In the present study, we calculated the AHE considering the FM state (i.e., with a large magnitude of the magnetic moment along $c-$direction) and obtained a negative value of $\sigma_{yx}^A$, which justifies our results.
Thus, according to our proposed model, the anomalous sign reversal of the anomalous Hall conductivity (AHC) in this compound is governed by both the magnitude of the magnetic moment and the underlying magnetic interaction.

\section*{Methods}

{\it Experimental details.}
MnPtGa single crystals were grown by the modified Bridgman method. High purity Mn, Pt and Ga were mixed in the stoichiometric ratio and put in a point-bottom alumina crucible that was baked overnight under vacuum at 1200$^\circ$C. After sealing the crucible under vacuum $(\sim 10^{-5}$ mBar) inside a quartz tube, it was heated to 1100$^\circ$C and slowly cooled to 800$^\circ$C inside a resistive heating furnace. Powder x-ray diffraction (XRD) was characterized by a PANalytical x-ray diffractometer, equipped with a monochromatic Cu-$K_\alpha$ ($\lambda = 1.5406$~\AA) x-ray source at room temperature. Single crystal x-ray diffraction (SXRD) data were collected on a four-circle Bruker diffractometer employing Mo-$K_\alpha$ radiation at around 300~K. Diffracted x-rays were detected by a Bruker CCD detector where the crystal-to-detector distance was 50~mm resulting in a resolution of the SXRD data of approximately $(\sin\theta/\lambda)_{max} = 0.802457$~\AA$^{-1}$. The diffracted intensity was collected on the detector during rotation of the crystal by 0.5$^\circ$. Obtained cell parameters are $a~=~b~=~4.334(8)$~\AA~and $c~=~5.583(9)$~\AA. The crystals were oriented using the Laue diffraction technique and cut using a spark erosion electric discharge machine (EDM). Magnetic measurements were performed in a SQUID-VSM magnetometer (Quantum Design, USA) and electrical measurements were done in physical property measurement system (PPMS, Quantum Design).

{\it SXRD analysis.}
To establish the correct structural model, single-crystal x-ray diffraction (SXRD) is performed at room temperature, where we have two refinements categorized into two models. In Model A, the structure was refined employing the high symmetry space group $P6_3/mmc$ and in Model B, $P3m1$ was used as shown in Tables (Table.~\ref{compare_model} and~\ref{compare_ref}). One can see that upon choosing the lower symmetry space group $P3m1$ (Model B) we have more degrees of freedom for the refinement, thereby increasing the number of parameters. However, the extra parameters now refined do not improve the model making it hard to justify to describe the model in the trigonal setting as opposed to the hexagonal setting $P6_3/mmc$. Moreover, in the trigonal setting, the $z$ coordinate (See Table supplementary) is no longer constrained by symmetry as it was in the hexagonal setting. Upon refining $z$, the standard uncertainty is quite large, thereby making it unreliable. Ergo, Model A is best suited to describe the structure in our case. 

The present crystals were synthesized using the Bridgman technique, similar to that presented in Ref.~\cite{Srivastava.Devi.2020}. However, we found that the compound has not crystallized in the $P3m1$ structure but rather favors $P6_3/mmc$, similar to that reported in ~\cite{Cooley.Bocarsly.2020}. Nevertheless, the origin of its polymorphism is yet to be explored.

{\it Computational details.}
The first-principles density functional theory (DFT) calculations were performed using the projector augmented-wave (PAW) potentials~\cite{blochl.94} implemented in 
the Vienna Ab initio Simulation Package ({\sc Vasp}) code~\cite{kresse.hafner.94,kresse.furthmuller.96,kresse.joubert.99}.
For the exchange-correlation energy, the generalized gradient approximation (GGA) in the Perdew, Burke, and Ernzerhof (PBE) parametrization was used~\cite{pardew.burke.96}.
The energy cutoff for the plane-wave expansion was set to $400$~eV.
Optimizations of the lattice constant and atom positions (in the presence of the spin--orbit coupling) were performed using $10 \times 10 \times 8$ {\bf k}--point grid, using the Monkhorst--Pack scheme~\cite{monkhorst.pack.76}.

Direct DFT calculations were used to calculation of tight binding model in the maximally localized Wannier orbitals~\cite{marzari.vanderbilt.97,souza.marzari.01,marzari.mostofi.12}, constructed by {\sc Wannier90}~\cite{pizzi.valerio.20} software.
We also perform calculations with the fixed values and directions of Mn magnetic moment ({\sc i\_constrained} tag).
Similar technique was earlier used to study of the doping effect on the AHE in thin ferromagnetic film Sr$_{1-x}$Ca$_{x}$RuO$_{3}$~\cite{mathieu.asamitsu.04}.

As a convergence condition of the optimization loop, we took the energy change below $10^{-6}$~eV and $10^{-8}$~eV for ionic and electronic degrees of freedom.
The obtained optimized structure were examined in context of the dynamical stability, via phonon spectrum calculation (calculations include the spin--orbit coupling).
We used the direct {\it Parlinski--Li--Kawazoe} method~\cite{parlinski.li.97}, implemented in the {\sc Phonopy} software~\cite{phonopy1,phonopy2}.
The crystal symmetry was analysed using {\sc FindSym}~\cite{stokes.hatch.05} and {\sc SpgLib}~\cite{togo.tanaka.18}, while the momentum space analysis was done using {\sc SeeK-path} tools~\cite{hinuma.pizzi.17}.

\bibliography{a_PtMnGa}

\begin{acknowledgments}

Some figures in this work were rendered using {\sc XCrySDen}~\cite{kokalj.99} software.
A.P. is grateful to Laboratoire de Physique des Solides in Orsay (CNRS, University Paris Saclay) for hospitality during a part of the
work on this project.
We kindly acknowledge support by National Science Centre (NCN, Poland) 
under Project No.~2021/43/B/ST3/02166 (A.P). %AMO

\end{acknowledgments}

\section*{Author contributions}
G.D. and A.T. conceived the project. G.D. and B.B.M grew the single crystals of MnPtGa. SXRD experiments are peroformed by A.L. and P.P.J.. S.R. performed the SXRD analysis. G.D. performed the magnetotransport measurements. A.P. performed the theoretical analysis and simulation. G.D., S.D., A.P. and A.T. jointly discussed the results and wrote the manuscript with contribution from all authors.

\section*{Competing interests}
The authors declare no competing interests

{\bf Correspondence} and requests for materials should be addressed to A.T.

%%%%%%%%%%%%%%%%%%%%%%%%%%%%%%%%%%%
%%%%%%%%%%%%%%%%%%%%%%%%%%%%%%%%%%%
%%%%%%%%%%%%%%%%%%%%%%%%%%%%%%%%%%%

\clearpage
\newpage

\onecolumngrid

\begin{center}
  \textbf{\Large Supplemental Material}\\[.3cm]
  \textbf{\large Unveiling the interplay of magnetic order and electronic band structure on the evolution of anomalous Hall effect in MnPtGa single crystal}\\[.3cm]
  %%%%%%
  Gourav Dwari {\it et al.}\\[.2cm]
  %%%%%%
  %{\itshape
%	$^{1}$Institute of Nuclear Physics, Polish Academy of Sciences, W. E. Radzikowskiego 152, PL-31342 Kraków, Poland
  %}
(Dated: \today)
\\[1cm]
\end{center}

\setcounter{equation}{0}
\renewcommand{\theequation}{SE\arabic{equation}}
\setcounter{figure}{0}
\renewcommand{\thefigure}{SF\arabic{figure}}
\setcounter{section}{0}
\renewcommand{\thesection}{SS\arabic{section}}
\setcounter{table}{0}
\renewcommand{\thetable}{ST\arabic{table}}
\setcounter{page}{1}

%%%%%%%%%%%%%%%%%%%%%%%%%%%%%%%%%%%
%%%%%%%%%%%%%%%%%%%%%%%%%%%%%%%%%%%
%%%%%%%%%%%%%%%%%%%%%%%%%%%%%%%%%%%

\begin{figure}[!ht]
    \centering
    \includegraphics[width=0.8\textwidth]{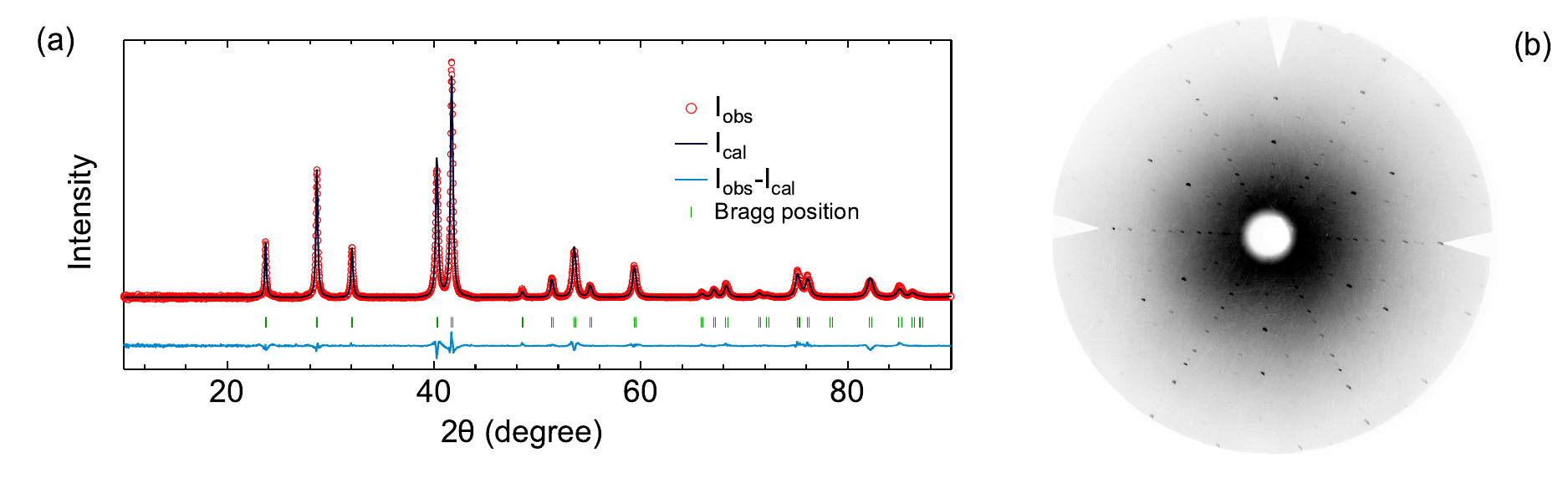}
    \caption{(a) Powder x-ray diffraction refined using Rietveld method with $P6_3/mmc$ space group at room temperature. (b) Laue diffraction pattern for (0001) plane at room temperature.}
    \label{XRD}
\end{figure}

\begin{figure}[!ht]
    \centering
    \includegraphics[width=0.35\textwidth]{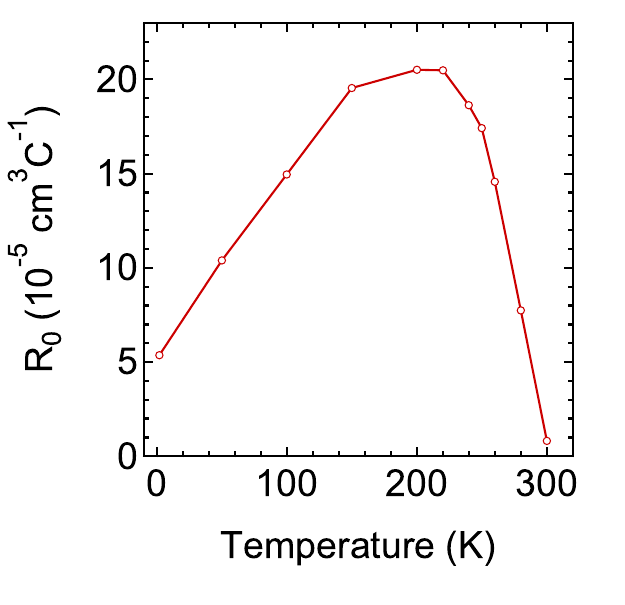}
    \caption{Temperature dependence of normal Hall coefficient $R_0$ exhibits no change in carrier type in the complete temperature range.}
    \label{NormalHallCoeff}
\end{figure}

\begin{figure}[!ht]
    \centering
    \includegraphics[width=1\textwidth]{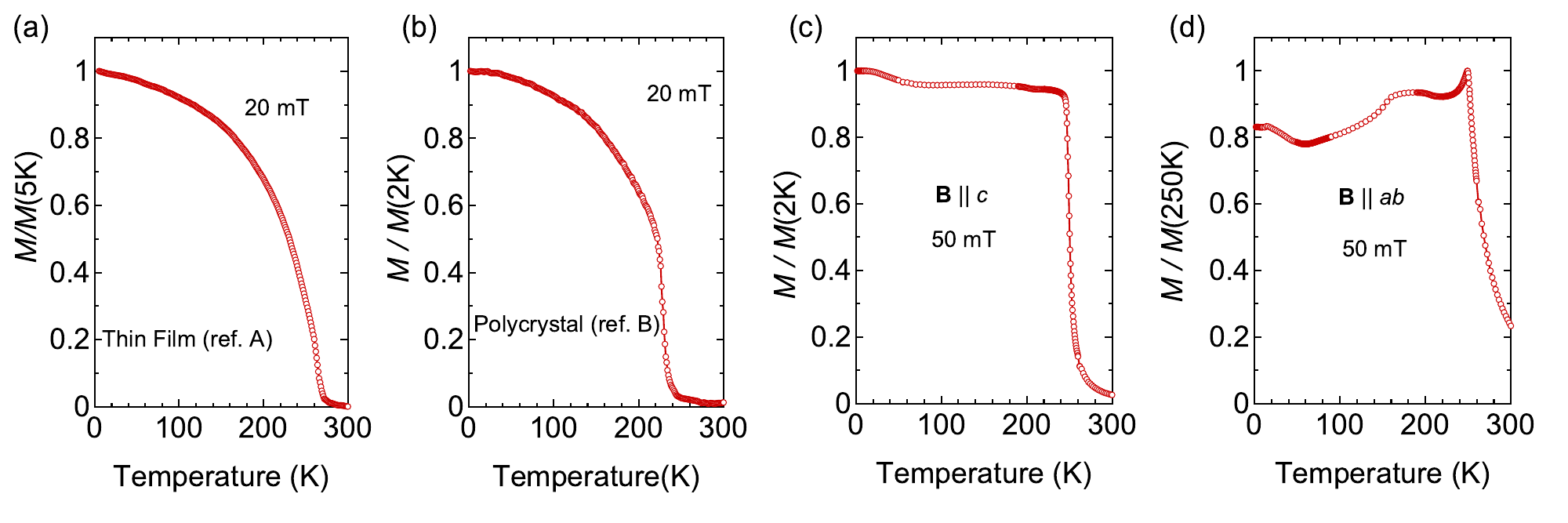}
    \caption{Comparision of magnetization (\textit{M}) of MnPtGa in (a) thin film (ref.A :~\cite{Ibarra.Lesne.2022}), (b) polycrystal (ref.B: ~\cite{Cooley.Bocarsly.2020}) and (c \& d) single crystal (this work) form.}
    \label{Magnzn_comparison}
\end{figure}

In this Supplemental Material, we present additional results:
\begin{itemize}
	\item Sec.~\ref{sec.sm.structure} Details of structural refinement\\
	 Table~\ref{compare_model} Structural refinement considering space group 194 and 156\\
	 Table~\ref{compare_ref} Details of refinement using spacegroup 194\\
	 Fig.~\ref{XRD} Room temperature powder XRD and back scattered Laue image along [0001] direction.
	
	\item Sec.~\ref{sec.sm.Hall}  Details of Hall resistivity and magnetism\\
	 Fig.~\ref{NormalHallCoeff}--Variation of $R_0$ (cf. eqn. \ref{eqn:Hall}) with temperature.\\
	 Fig.~\ref{Magnzn_comparison} Comparison of magnetization data with literature.
	
	\item Sec.~\ref{sec.sm_theo}  Theoretically obtained crystal structure and electronic structure\\
	 Fig.~\ref{fig.cryst} Phonon dispersion curves with magnetic moments along $c-$axis and tilted from $c-$axis with space group $P6_3/mmc$.\\
	 Fig.~\ref{fig.colinear} Comparison of spin-dependent electronic band structure.
\end{itemize}

\section{Details of structural refinement}
\label{sec.sm.structure}
\begin{table}[!ht]
\caption{\label{compare_model}%
Results of indexing and integration of the diffraction
data of PtMnGa at $T = 293$ K, assuming
alternatively hexagonal and trigonal symmetries.}
\centering
\begin{ruledtabular}
\begin{tabular}{ccc}
Model & A & B\\
Crystal system & Hexagonal & Trigonal \\
Space group & $P6_3/mmc$ (No.~194) & $P3m1$ (No.~156) \\
$R_{exp}$ (obs/all) (\%)&1.61/1.62 &1.94/1.97 \\
$R_{int}$ (obs/all) (\%) &5.00/5.06 &6.12/6.22 \\
Unique refl (obs/all) & 87/95 &142/170 \\
$R_{F}$ (obs/all) (\%) &1.37/1.66 &1.41/2.03 \\
No. of parameters& 8 & 20 \\
\end{tabular}
\end{ruledtabular}
\end{table}

\begin{table*}[ht]
\centering
\scriptsize
\caption{\label{compare_ref}Refinement comparison of models A and B to
see which space group yields the best possible fit. The criterion of observability: $I>3\sigma(I)$.
Given are the fractional coordinates $x$, $y$, $z$ of the atoms,
their anisotropic displacement parameters (ADPs)
$U_{ij}$ $(i, j = 1, 2, 3)$ and the equivalent isotropic displacement parameter $U^{eq}_{iso}$.
Refinement method used: Least squares on $F$.}
\begin{ruledtabular}
\begin{tabular}{ccccccccccccc}
Atom & $x$ & $y$ & $z$ & $U_{11}$ & $U_{22}$ & $U_{33}$ & $U_{12}$ & $U_{13}$ & $U_{23}$ & $U^{eq}_{iso}$ \\
\hline
\multicolumn{4}{l}{Space group: $P6_3/mmc$} & & & & & & & \\
Mn &0 &0 &0 & 0.0094(4) & 0.0094 & 0.0074(7) & 0.0047 & 0 & 0 & 0.0087(3) \\
Pt &2/3 & 1/3 & 1/4 & 0.0071(2) & 0.0071 & 0.0125(3) & 0.0036 & 0 & 0 & 0.0089(1) \\
Ga &1/3 &2/3 &1/4 & 0.0066(4) & 0.0066 & 0.0162(6) &0.0033 &0 &0 & 0.0098(3) \\
\hline
\multicolumn{4}{l}{Space group: $P3m1$} & & & & & & & \\
Mn1 & 0 & 0 & 0.0056(47) & 0.0096(5) & 0.0096 & 0.0080(11) & 0.0048 & 0 & 0 & 0.0091(4) \\
Mn2 &0 &0 &0.5052(46) & 0.0098(4) & 0.0098 & 0.0079(9) & 0.0049 & 0 & 0 & 0.0092(4) \\
Pt1 &1/3 &2/3 &0.2480(7) & 0.0068(6) & 0.0068 & 0.0116(11) &0.0034 &0 &0 & 0.0084(5) \\
Pt2 &2/3 &1/3 &0.7482(6) & 0.0075(6) & 0.0075 & 0.0149(13) &0.0038 &0 &0 & 0.0099(5) \\
Ga1 &1/3 & 2/3 &0.7593(33) & 0.0114 & 0.0114 & 0.0134(37) &0.0057 &0 &0 & 0.0012(15) \\
Ga2 & 2/3 &1/3 &0.2577(29) & 0.033(15) & 0.0033 & 0.0154(33) & 0.0017 &0 &0 & 0.0074(13) \\
\end{tabular}
\end{ruledtabular}
\end{table*}

\vspace{2cm}

\section{Details of Hall resistivity  and magnetism}
\label{sec.sm.Hall}


\newpage

\section{Theoretically obtained crystal structure}
\label{sec.sm_theo}

\begin{figure*}[!ht]
\includegraphics[width=\textwidth]{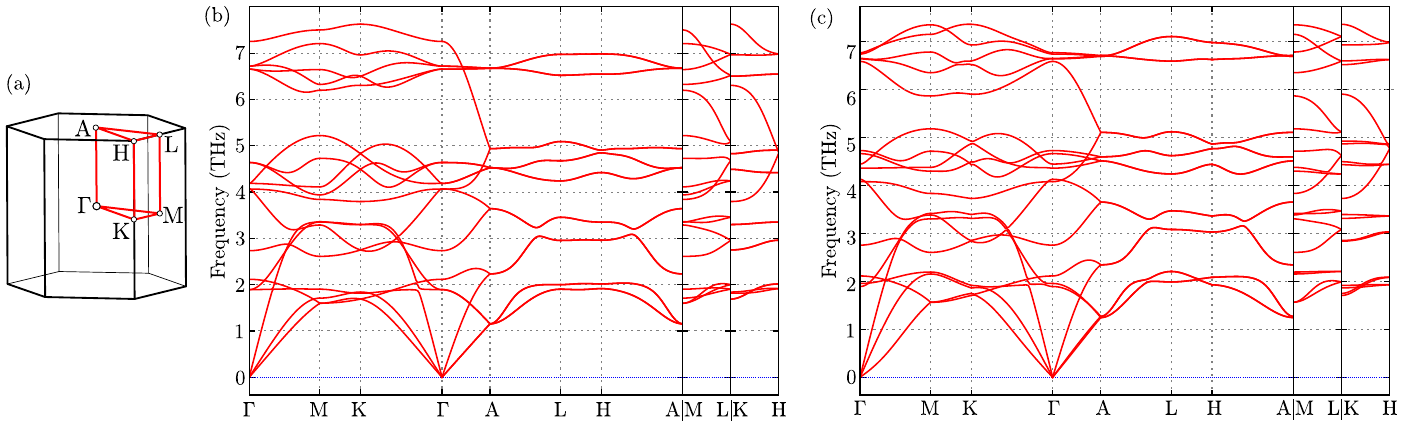}
\caption{The phonon dispersion curves for system with magnetic moments along $c$-direction (b) and titles from $c$-direction (c).
To present results for both structures, we used the $P6_{3}/mmc$ Brillouin zone, presented on panel (a).
For the magnetic moments along $c$-direction, the system contain $P6_{3}/mmc$ symmetry. 
Contrary to this, tilted of the magnetic moments from $c$-direction, lead to the symmetry breaking and realization of $P2_{1}/m$ symmetry.
}
\label{fig.cryst} 
\end{figure*}

From ``full'' optimization of the system within DFT calculations we get two structures, depending on the Mn-magnetic moment orientation:
\begin{itemize}
\item for Mn magnetic moments along $c$-direction, the system possesses hexagonal $P6_3/mmc$ structure (space group No.~194) with lattice parameters $a_{h} = b_{h} = 4.345$~\AA, $c = 5.526$~\AA, and $\alpha = \beta = 90^{\circ}$, $\gamma = 120^{\circ}$. 
This corresponds to the volume of $90.34$~\AA$^{3}$.
Atom positions were found as Mn $2a$ (0,0,0), Ga $2c$ (1/3,2/3,1/4), and Pt $2d$ (1/3,2/3,3/4).
%%%%%%%%%%%%%%%%%%%%%%%%%%
\item for Mn magnetic moment tilted from $c$-direction, the system possesses monoclinic $P2_1/m$ structure (space group No.~11) with lattice parameters  $a_{m} = 4.339$~\AA, $b_{m} = 5.509$~\AA,  $c_{m} = 4.347$~\AA, and, $\alpha = 90^{\circ}$, $\beta = 119.868^{\circ}$, $\gamma = 90^{\circ}$.
This corresponds to the volume $90.11$~\AA$^{3}$.
Atom positions were found as Mn $2a$ (0,0,0), Ga $2e$ (1/3,1/4,0.66696), and Pt $2e$ (0.66660,1/4,0.33284).
We estimate the angle of the Mn magnetic moment from $c$-direction as $39.88^{\circ}$.
\end{itemize}
Experimentally obtained lattice constant $a_{h} = b_{h} = 4.3348$~\AA, and $c_{h} = 5.5839$~\AA ($V = 90.86$~\AA$^3$) are in good agreement with previously reported at $300$~K, for $P6_{3}/mmc$ ( as $a_{h} = b_{h} = 4.330$~\AA, and $c = 5.573$~\AA, $V = 90.50$~\AA$^{3}$~\cite{Cooley.Bocarsly.2020}).
Additionally, theoretical results are in agreement with those obtained theoretically.

First, we should notice, that the basis for the obtained monoclinic phase is rotated with respect to the hexagonal structure, i.e. $a_{h} \rightarrow a_{m}$, $b_{h} \rightarrow c_{m}$, and $c_{h} \rightarrow b_{m}$.
From this, we can see that the lattice distortion introduced by the magnetic moment tilted is relatively small, while lattice parameters are close to the reported experimentally.
The system structure modification comes from the shift of the atoms from the high symmetry position of the hexagonal structure due to the magnetic moment tilting and system symmetry breaking.

Second, we should remember that MnPtGa at low temperatures contain a long-range magnetic order, with a periodicity of around 13 lattice units (at $50$~K and $Amm$ symmetry), or around 12 lattice units (at $10$~K and $C2/m$ symmetry)~\cite{Cooley.Bocarsly.2020}.
This corresponds to a relatively large magnetic unit cell, with the longest lattice constant around $70$~\AA.
Nevertheless, in such cases, the related unit cell containing one chemical formula (and corresponding for $P6_{3}/mmc$ symmetry) has $c_{h} = 5.472$~\AA\ at $50$~K, or $c_{h} = 5.642$~\AA\ at $10$~K.
These values are close to this obtained theoretically for both structures, i.e. $c_{h}$ for $P6_3/mmc$, and $b_{m}$ for $P2_1/m$.

Finally, both structures obtained for magnetic moment along $c$-direction and tilted from $c$-direction are stable in a dynamical sense. 
The phonon dispersion curves do not possess the imaginary soft modes (Fig.~\ref{fig.cryst}) -- results were obtained in the presence of the spin-orbit coupling.
Nevertheless, tilting of the magnetic moments from $c$-direction significantly affects the electronic band structure (see Fig.~\ref{fig.theo} in the main text).

\begin{figure*}[!ht]
\includegraphics[width=0.7\textwidth]{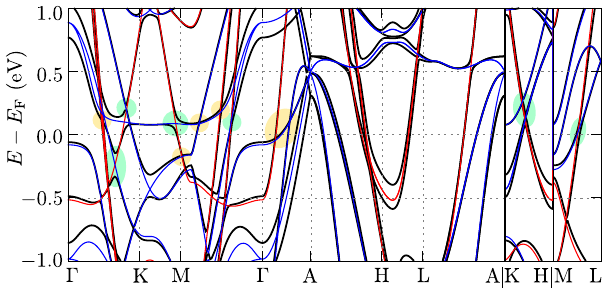}
\caption{
Comparision of the spin-dependent electronic band structure (blue and red for spin-up and spin-down, respectively) with the band structure obtained in the presence of spin--orbit coupling (black lines).
Results for magnetic moments along $c$-direction.
Introduction of the spin--orbit coupling opens several gaps between bands (green-marked regions).
However, the few band crossings are still preserved close to the Fermi level (yellow-marked regions).
}
\label{fig.colinear} 
\end{figure*}

%\bibliography{PtMnGa}

\end{document}